# POINT TRANSFORMER FOR CORONARY ARTERY LABELING

*Xu Wang, Jun Ma, Jing Li\**

Shen Zhen Raysight Intelligent Medical Technology Co., Ltd., Shenzhen, China

**ABSTRACT**

Coronary CT angiography (CCTA) scans are widely used for diagnosis of coronary artery diseases. An accurate and automatic vessel labeling algorithm for CCTA analysis can significantly improve the diagnostic efficiency and reduce the clinicians' manual efforts. In this paper, we propose a simple vessel labeling method based on the Point Transformer [7], which only needs the coronary artery segmentation. Specifically, firstly, the coronary segmentation is transformed to point cloud. Then, these points are fed into the hierarchical transformer blocks to obtain the multi-level features, including local and global features. Finally, the network output the semantic classification points and map them to centerline labeling. This method is only based on the structure of coronary segmentation and need not other features, so it is easy to generalize to other vessel labeling tasks, e.g., head and neck vessel labeling. To evaluate the performance of our proposed method, CCTA scans of 53 subjects are collected in our experiment. The experimental results demonstrate the efficacy of this approach.

## 1. INTRODUCTION

Cardiovascular diseases are the leading cause of death globally, taking an estimated 17.9 million lives each year [1]. CCTA as a front-line imaging for diagnosis of coronary artery diseases has been widely used by radiologists. In clinic standard workflow, three steps are usually performed: 1) segmenting the coronary tree by semi-auto software; 2) labeling all vessels manually; 3) reconstructing the CPR (curved planar reformation) images and diagnosing the stenosis and plaque type of the main coronary arteries. As coronary tree structure is complicated (see Fig. 1), vessels labeling is a challenging task in diagnosis workflow. Therefore, an automatic and accurate coronary artery labeling technique can greatly improve the diagnostic efficiency and reduce clinicians' manual efforts.

Recently, many related studies have been presented to tackle the vessel labeling challenge. To the best of our knowledge, the published methods can be roughly divided into two categories: matching-based method and learning-based method.

The matching-based methods were developed to match the common structure of coronary tree. Yang, et al. [2] proposed a two-step matching algorithm based on the mean template of a statistical coronary tree model. Zhang, et al. [3] combined the geometry structure of the coronary tree and the heart to label the vessels. Although there are some expert consensus guidelines for labeling coronary vessels, it is difficult to find a standard matching template to handle some complex cases as the coronary tree of patients are complicated and varied.

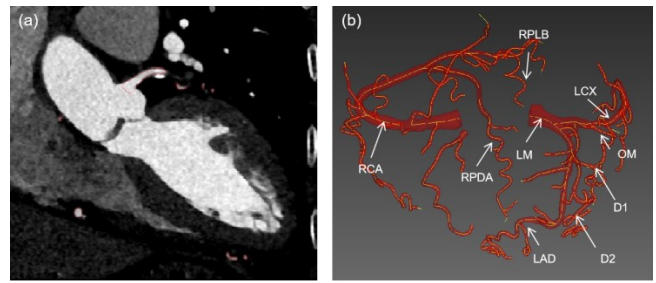

**Fig. 1.** An example of coronary CTA image. (a) A coronal slice of CTA; (b) annotated segments: LM、RCA、LAD、LCX、OM、D1、D2、RPLB、RPDA.

The learning-based methods were based on the deep learning or machine learning method to predict vessel labeling. Specifically, Wu, et al. [4] developed a bi-directional tree structure LSTM algorithm to predict vessel labeling. Yang, et al. [5] designed a CPR-GCN algorithm. Each branch was abstracted as the higher-level features. Branch features were extracted along centerlines, and then were fed to the bi-directional LSTM network. In recent years, point cloud-based method has been becoming more and more popular. Li et al. [6] classified each voxel in the coronary segmentation. In this study, they combined the points cloud from the coronary tree and cardiac mask surfaces so as to provide more local and global information. Yao, et al. [12] labeled head and neck vessels based on point cloud network [8] and graph convolutional network [14]. However, these point cloud-based approaches were insufficient to employ the local relationship with the neighbor points, which is useful to distinguish the bifurcation branches labeling or lumen diameter.

In this paper, we propose a simple and accurate vessel labeling method based on the Point Transformer [7] which only need the coronary segmentation as input. In clinic, radiologists can use the coronary tree alone to identify all the vessel labeling, so we imitate radiologist behavior by using the Point Transformer to extract not only the global features such as the coronary geometry structure but also the local features which represent the branch bifurcation or lumen diameter.

Our contributions of the paper can be summarized as follows:

(1) we propose a novel architecture inspired by point-transformer [7]. As far as we know, it is the first application in coronary vessel labeling based on the Point Transformer method.

(2) the advantage of Point Transformer method to utilize not only the local features between the neighbor points [10], but also the global feature such as coronary geometry structure.

(3) our method achieved an outstanding performance on a dataset consisting of 53 CCTA images.

## 2. METHOD

Our method consists of two stages: 1) As shown in Fig. 2, we firstly transform the foreground voxel-level segmentation to point cloud and then feed it into the hierarchical Point Transformer blocks to get the semantic classification points; 2) we then map the labeling from voxel-level to centerline-level.

### 2.1. Voxel-level Labeling by Point Transformer

A vessel point cloud can be transformed from the binary coronary vessel segmentation S which is annotated by radiologists. Note that only the foreground points of the S are extracted with M points. We denote the point cloud P as P = [$p_1$, $p_2$, ..., $p_N$] with N points sampled in M. The point is set as $p_i$ = (x, y, z, x', y', z') where (x, y, z) is the point position and the (x', y', z') is the point position after normalization, which avoids different origin of CCTA scans. Therefore, the constructed point cloud can be denoted as a matrix of size 6 × N.

As shown in Fig. 2, the network is composed of encoder and decoder parts like U-Net [9], which is inspired by the Point Transformer [7] and Local relation network [10]. First, in the encoder part, we input the sampled point cloud to 2*(Conv+BN+ReLU) with convolution kernel size 1 and output channel 32. Then we get the feature maps with size 32 × N and feed them into Transformer part which contains several the cascaded Point Transformer block modules and TransitionDown modules. Similarly, the decoder consists of several TransitionUp modules and the cascaded Point Transformer block modules. We use the structures and parameters as same as the Point Transformer [7] for the Point Transformer block, TransitionDown and T-

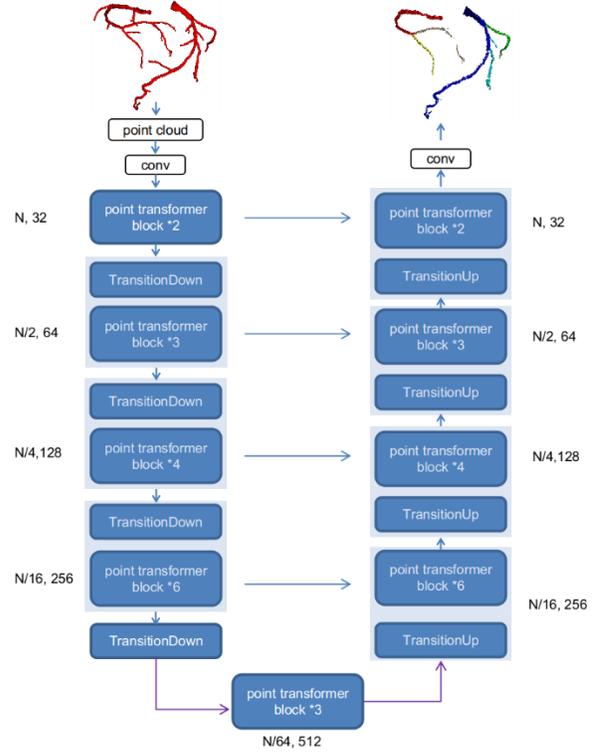

**Fig. 2.** Framework of the vessel labeling transformer. Where N is the number of input points, [32, 64, 128, 256, 512] is the number of channels in each stage.

ransitionUp modules. Finally, we use a convolution (with kernel size 1 and output channel K) to get the score map with size K×N and apply softmax+argmax to output the final classification. Besides, to keep the representation of small vessels, we use the skip connection between the encoder and decoder parts to integrate the shallow and the deep features as shown in Fig. 2. The Point Transformer block consists of a transformer layer and 2 linear layers. In order to precisely label the bifurcation of coronary tree, we use the transformer layer to learn the anatomical spatial relationship between the local point and its neighbor points [10]. The next layer feature of the point can be calculated by the set of points in the local neighborhood based on the self-attention network:

$$y_i = \sum_{x_j \in x} \rho(\gamma(k(x_i) - q(x_j) + \delta)) \odot (v(x_j) + \delta) \qquad (1)$$

where $x$ is the feature map of neighbor $i$, $\rho$ is a softmax operation. $x_i$ is input feature, $y_i$ is output feature and $\gamma$, $k$, $q$, $v$ is a linear transformation, $\delta$ is the learnable relative encoding of position, and the positional relationship between $i$ and $j$. Specifically, we calculate the attention of the nearest h (set as 16 empirically) neighbors and sum these attention values. The local and global features can be learned in the shallow and deep layers, respectively. Benefiting from the point cloud construction, the transform-

**Table 1.** Comparison of previous work with ours in branch-level centerline labeling.

| Method (%) | Yang [2] | Cao [11] | Wu [4] | Li [6] | Ours |
|---|---|---|---|---|---|
| LM | 99.3 | **100.0** | 99.1 | 95.7 | **100.0** |
| LAD(p/m/d) | 93.4/86.8/93.4 | 93.6/85.8/95.4 | 96.9 | **99.1** | 99.03 |
| LCX(p/m/d) | 84.6/80.3 | 87.3/83.2 | 93.5 | **95.1** | 94.2 |
| RCA(p/m/d) | 97.8/94.1/92.7 | 85.1/82.3/92.5 | 96.0 | 97.6 | **98.2** |
| D(1/2) | 100/86.8 | 93.5/82.2 | 91.0 | 86.5 | **93.4** |
| OM(1/2) | 86.1/78.8 | 90.4/79.7 | 85.2 | 82.9 | **86.5** |
| PLB(R/L) | 88.3/ - | 89.8/85.7 | 82.7/65.9 | 95.5 | **98.6/-** |
| PDA(R/L) | 94.1 | 96.6 | 79.8 | 89.6 | **97.6/-** |

*(p/m/d) denote coronary artery proximal segment, middle segment and distal segment.

er layer is suitable to get not only the anatomical geometry of the coronary tree, but also the local information of the bifurcations or lumen diameter.

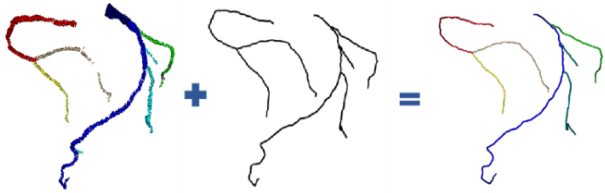

**Fig. 3.** Centerline-level labeling.

### 2.2. Mapping voxel-level to centerline-level labeling

As voxel-level labeling may come with noises or corruptions, we map the voxel-level labeling to centerline-level labeling, which is useful to develop downstream processing, such as stenosis degree and plaque type analysis. We generate the centerline-level labeling based on the highest overlap rate between the voxel-level segmentation and the dilated centerline. Specifically, we first dilated the centerline which is extracted by radiologists to tubular voxel mask with radius 1mm. Then we compute each centerline's overlap rate which is defined as:

$$O_n = \underset{m}{\mathrm{argmax}} \frac{L_m \cap S_n}{L_m} \quad (2)$$

where $L_m$ is the unlabeled dilated centerline $m$, $S_n$ is the voxel-level labeling segmentation of category $n$, and $O_n$ is the overlap rate of centerline $m$. Finally, we label each centerline based on its highest overlap rate as presented in Fig. 3.

## 3. EXPERIMENTS AND RESULTS

### 3.1. Data and Implementation

We collected 373 CCTA scan with or without Nitroglycerin, which were approved and consented by the ethics committee. The mean spacing of these CCTA scans was 0.35 * 0.35 * 0.60 mm³. In addition, these scans were annotated by two experienced physicians. The ground truth consisted of the labeled centerlines, the coronary tree segmentation and its corresponding voxel-level label. Labeling of the centerline was including LM, LAD, LCX, RCA, D, OM, PLB and PDA. We used 320 scans for training and 53 for testing. In the training phase, we did the online data augmentation such as rotation and translation. We used a RTX 3090 GPU with 4 batch size and implemented the algorithm with PyTorch [13] framework. Our model was trained with 120 epochs, using SGD optimizer with learning rate 0.001 and decay 10-4. We sampled N=12288 points and fed them into the network. Our network had 5 levels and the downsampling rates were 2, 2, 4, 4 respectively.

**Table 2**. Comparison between PointNet and our method in voxel-level coronary labeling.

| Method | Precision (%) | Recall (%) | F1-score (%) |
|---|---|---|---|
| PointNet [8] | 73.8 | 71.7 | 69.9 |
| Ours | **88.3** | **87.9** | **87.1** |

### 3.2. Quantitative Result

In Table 1, we compared our results with 4 other existing methods and the evaluation was based on the branch-level centerline. The metric is precision of the branch-level centerline labeling as proposed in [11]. As shown in Table 1, precision of the main 4 segments (LM, LAD, LCX and RCA) were all >94%, and precision of all other segments are all >85%. OM segment has the lowest precision 86.5%, due to that fact that in some patients OM2 vessels are longer and thicker than distal LCX and this confuses the algorithm.

As coronary segmentation labeling is a significant step, we compared our method with PointNet that is a popular point-cloud method. As shown in Table 2, the performance of our method based on the Point Transformer is better than PointNet.

## 3.3. Ablation Experiments

In the stage of converting sampling points to point cloud, the number of sampled points would affect the voxel-level labeling results. Therefore, we compared the mean accuracy of voxel classification for different point numbers in the testing data. As shown in Table 3, the accuracy increases from 92.64% (4096 points) to 93.12% (12288 points), then decreases to 93.04% (16384 points). Fig. 4 shows example results from different settings of sampling points. In Fig. 4(c) the result with 8192 points loses R-PLB branch information, and in Fig.4(e) result with 16384 points yields wrong classifications. Therefore, we used 12288 sampling points in our Point Transformer network.

**Table 3.** Result with different sampling point number.

| Point Number | 4096 | 8192 | **12288** | 16384 |
|---|---|---|---|---|
| Accuracy (%) | 92.64 | 92.89 | **93.12** | 93.04 |

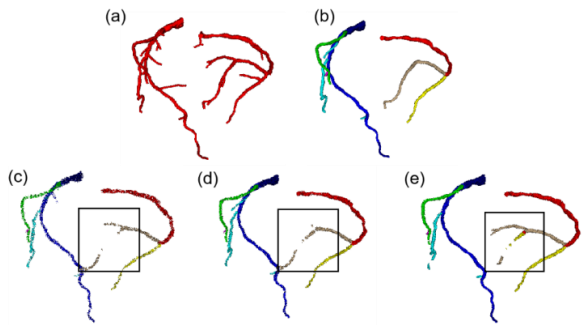

**Fig. 4.** Results with different setting of sampling points. (a) The original coronary segmentation; (b) The corresponding voxel-level labeling; (c) (d) (e) Results with 8192, 12288 and 16384 sampling points respectively.

## 4. CONCLUSION

In this paper, we present a novel vessel labeling method based on the Point Transformer. Our method only uses the segmentation of coronary tree to predict the vessels labeling, which is similar to radiologist behavior. As this method only based on the geometry structure of coronary tree, it is easy to generalize to other vessel labeling tasks. This method incorporates both the coronary tree structure and spatial information of all major and minor vessels. To evaluate the performance of our proposed method, 53 CCTA scans are used in our experiment. The experimental results demonstrate the efficacy of the present approach. In future work, we will combine our centerline labeling method based on transformer with automatic extraction of coronary segmentation and centerline, which will achieve full automation in inference process.

## 5. ACKNOWLEDGMENTS

This work was supported by National Key R&D Program of China (grand No. 2022YFC2409000).

## 6. REFERENCES

[1] Read, S.H. and Wild, S.H., 2020. Prevention of premature cardiovascular death worldwide. The Lancet, 395(10226), pp.758-760.

[2] Yang G, Broersen A, Petr R, et al. Automatic coronary artery tree labeling in coronary computed tomographic angiography datasets[C]//2011 Computing in Cardiology. IEEE, 2011: 109-112.

[3] Zhang C J, Xia D, Zheng C, et al. Automatic identification of coronary arteries in coronary computed tomographic angiography[J]. IEEE Access, 2020, 8: 65566-65572.

[4] Wu D, Wang X, Bai J, et al. Automated anatomical labeling of coronary arteries via bidirectional tree LSTMs[J]. International journal of computer assisted radiology and surgery, 2019, 14(2): 271-280.

[5] Yang H, Zhen X, Chi Y, et al. Cpr-gcn: Conditional partial-residual graph convolutional network in automated anatomical labeling of coronary arteries[C]//Proceedings of the IEEE/CVF Conference on Computer Vision and Pattern Recognition. 2020: 3803-3811.

[6] Li Z, Xia Q, Wang W, et al. Segmentation to Label: Automatic Coronary Artery Labeling from Mask Parcellation[C]//International Workshop on Machine Learning in Medical Imaging. Springer, Cham, 2020: 130-138.

[7] Zhao H, Jiang L, Jia J, et al. Point transformer[C]//Proceedings of the IEEE/CVF International Conference on Computer Vision. 2021: 16259-16268.

[8] Qi C R, Su H, Mo K, et al. Pointnet: Deep learning on point sets for 3d classification and segmentation[C]//Proceedings of the IEEE conference on computer vision and pattern recognition. 2017: 652-660.

[9] Ronneberger O, Fischer P, Brox T. U-net: Convolutional networks for biomedical image segmentation[C]//International Conference on Medical image computing and computer-assisted intervention. Springer, Cham, 2015: 234-241.

[10] Hu H, Zhang Z, Xie Z, et al. Local relation networks for image recognition[C]//Proceedings of the IEEE/CVF International Conference on Computer Vision. 2019: 3464-3473.

[11] Cao Q, Broersen A, de Graaf M A, et al. Automatic identification of coronary tree anatomy in coronary computed tomography angiography[J]. The international journal of cardiovascular imaging, 2017, 33(11): 1809-1819.

[12] Yao L, Jiang P, Xue Z, et al. Graph convolutional network based point cloud for head and neck vessel labeling[C]//International Workshop on Machine Learning in Medical Imaging. Springer, Cham, 2020: 474-483.

[13] Paszke A, Gross S, Massa F, et al. Pytorch: An imperative style, high-performance deep learning library[J]. Advances in neural information processing systems, 2019, 32.

[14] Wolterink J M, Leiner T, Išgum I. Graph convolutional networks for coronary artery segmentation in cardiac CT angiography[C]//International Workshop on Graph Learning in Medical Imaging. Springer, Cham, 2019: 62-69.